# Publication patterns of award-winning forest scientists and implications for the ERA journal ranking

*Jerome K Vanclay*
Southern Cross University
PO Box 157, Lismore NSW 2480, Australia
Tel +61 2 6620 3147, Fax +61 2 6621 2669, JVanclay@scu.edu.au

**Abstract**

Publication patterns of 79 forest scientists awarded major international forestry prizes during 1990-2010 were compared with the journal classification and ranking promoted as part of the 'Excellence in Research for Australia' (ERA) by the Australian Research Council. The data revealed that these scientists exhibited an elite publication performance during the decade before and two decades following their first major award. An analysis of their 1703 articles in 431 journals revealed substantial differences between the journal choices of these elite scientists and the ERA classification and ranking of journals. Implications from these findings are that additional cross-classifications should be added for many journals, and there should be an adjustment to the ranking of several journals relevant to the ERA Field of Research classified as 0705 Forestry Sciences.

**Graphical abstract**

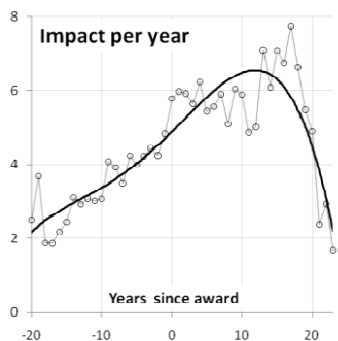

**Highlights**

1. Prize-winners exhibit elite performance a decade before and two decades following their award;
2. Their publication patterns correlate with other established metrics but conflict with the ERA classification and ranking;
3. Their publication choices suggest that the ERA ranking warrants revision and that several forestry journals including *Forest Ecology and Management, Tree Physiology* and *Canadian Journal of Forest Research* warrant an A* ranking within ERA;
4. Multiple fields of research should be assigned within the ERA classification to recognise and encourage interdisciplinary collaboration.

**Introduction**

Australia recently commenced an initiative called Excellence in Research for Australia (ERA) to monitor and stimulate university research performance (Anon, 2009; Hicks, 2009). Several indicators are used to rank institutional performance in pre-defined research fields, encouraging the formation of research teams aligned with the defined fields. Principal among the indicators is a four-tiered journal ranking that serves as a proxy for the quality of research outputs. Thus the ranking of journals and the definition of research fields will be influential in shaping future university research in Australia, and warrants careful scrutiny (Lamp, 2009; Bloch, 2010; De Lange et al, 2010; Lamp & Fisher, 2010; Vanclay, 2011).

ERA assesses research outputs by Australian universities within defined Fields of Research (FORs) defined by the Australian and New Zealand Standard Research Classification (ANZSRC, 2008). The ANZSRC is a hierarchical classification that involves a 2-digit division (e.g., 08 Information and Computing Sciences), a 4-digit group (0807 Library and Information Studies) and a 6-digit FOR (080705 Informetrics). The 4-digit research groups are used by ERA both to monitor university performance and to classify and rank journals as indicators of performance. Of 20 712 journals recognised by ERA, 13 836 are assigned to a single FOR, 6 273 are assigned to two or more FORs (either through allocation to two or three FOR groups, or by allocation to one or more FOR divisions), and 603 journals are denoted multidisciplinary.

Within each FOR, journals are ranked into four categories, A*, A, B and C, nominally representing 5, 15, 30 and 50 percentiles so that A* should represent the top 5% of journals, A should include the next 15%, B the next 30%, and C the remaining 50% of journals (Graham, 2008). The journal ranking proposed by the ERA has been controversial (e.g., Peters, 2008; Haslam and Koval, 2010), and forestry is one of the fields of research where the journal ranking appears deficient (Vanclay, 2011). Similar criticism has been attracted by the Research Assessment Exercise (Bence & Oppenheim, 2004) and Research Excellence Framework in the United Kingdom (Johnston, 2009). The ERA has chosen to use a subjective expert ranking of journals, and it is appropriate that such rankings should be compared against other quantitative approaches, notwithstanding limitations of ranking systems (Stringer et al., 2008; Lawrence, 2008).

Deficiencies in the ERA ranking, especially within the fields of agriculture (i.e., 07 Agricultural and Veterinary Sciences) and forestry (0705 Forestry Sciences) have previously been identified (Vanclay, 2011), so it is appropriate to scrutinize these deficiencies to indicate adjustments to the classification and rankings. Other researchers have examined rankings based on journal citations (e.g., Thomas and Watkins, 1998; Vanclay, 2008a,b; Bontis and Serenko, 2009; Harzing and van der Wal, 2009; Moed, 2010), and the present study seeks to offer complementary evidence based on the publication patterns of prominent forest scientists. The study was confined to publications and citations seen by Scopus, the official data provider to ERA in 2010, but other researchers have examined the similarity between Scopus and other citation providers (e.g., Falagas et al, 2008; Bollen et al, 2009; Li et al, 2010; Rocha-e-Silva, 2010; Siebelt et al, 2010; Hall, 2011).

The assumption underlying the following tests is that recipients of prestigious prizes are experienced scientists who are likely to be discerning in their choice of publication outlet, and who are likely to choose journals of good quality and wide reach, attributes that should be reflected in the ERA classification and ranking. Hence, the present study examines publication patterns of recipients of four major international prizes for scientific achievement in forestry, the Scientific Achievement Award of the International Union of Forest Research Organizations (IUFRO), the Marcus Wallenberg Prize, the Queen's Award for Forestry, and the Schweighofer Prize.

**The Forestry Prizes**

The IUFRO Scientific Achievement Award has been presented at each IUFRO World Congress (approximately once every 5 years) since 1971, to recognise the greatest achievement in each of several (5 awards in 1971, increasing to 11 in 2010) subject areas within IUFRO (currently silviculture, physiology and genetics, forest operations, forest assessment, modelling and management, forest products, forest health, forest environment, social sciences, forest policy and economics). These awards are made in recognition of "research results published in scientific journals, proceedings of scientific meetings or books, appropriate patents or other relevant evidence that clearly demonstrates the importance of the scientific or technical achievement to the advancement of regional or world forestry or forest research" (Anon, 2011a). To date, 77 scientists from 26 different countries have been honoured with this award (Anon, 2011b).

The Marcus Wallenberg Prize has been awarded annually since 1981. The purpose of this Prize is "to recognize, encourage and stimulate path-breaking scientific achievements which contribute significantly to broadening knowledge and to technical development within the fields of importance to forestry and forest industries" (Anon, 2011c). The Prize may be awarded to individuals or to groups of up to 4 researchers, and to date, 47 individuals from 7 countries have been recognised, either as individuals or team members. The prize recognises achievements across the breadth of the forestry sector, including both field forestry (genetics, systematics and tree breeding; silviculture and agroforestry; forest ecology and tree physiology; biometrics, computing and remote sensing; forest management, forest protection; forestry operations) and forest products (wood and wood processing; papermaking fibres; paper- and board-making processes; recycling of forest products; innovations to improve wood use and environmental performance).

The Queen's Award for Forestry recognises outstanding contributions to forestry by an outstanding mid-career forester who "combines exceptional contributions to forestry with an innovative approach to his or her work" (Anon, 2011d). The Queen's Award for Forestry is not confined to researchers, and has recognised other achievements of awardees with few publications. The award has been made nine times since 1987 to foresters from Australia, India, Malaysia, United Kingdom, and Zimbabwe.

The Schweighofer Prize recognises "innovative ideas, technologies, products and services concerning the whole value chain in order to strengthen the competitiveness of the European forest-based sector" (Anon, 2011e). The Schweighofer Prize has been offered every second year since 2003, with a total of four individuals (from Finland, Germany and Switzerland) receiving the main prize. Although the prize also includes several innovation prizes, the present analysis considers only the publication outputs of the main prize recipients.

**Materials and Methods**

The four prizes have been awarded a total of 137 times, but six individuals have received more than one of these prizes, so there are a total of 131 individuals who have been awarded one or more of these prizes. All but 14 of these individuals have publications visible in Scopus, the official data provider to ERA in 2010. Collectively, these 117 individuals created 6058 publications seen by Scopus. Standard citation data for all 6058 publications were exported from Scopus on 14 February 2011 in CSV format for further analysis. These publications included a wide range of material including conference proceedings, editorials, obituaries, and other minor contributions which were removed to leave 5518 contributions (articles and reviews) in 859 journals during the period 1958 to 2011 (but only 446 journals have >1 article). It is somewhat problematic surveying such a 54-year period because citation coverage is not uniformly thorough throughout, and because some journals ceased and others commenced during the period. Nonetheless, this collection of scientific output offers some interesting insights into contemporary publishing and citation patterns of prominent forestry scientists.

This study seeks to establish the time-frame over which prize-winning forest scientists may be regarded as 'elite', and contrasts their publication patterns during this elite period with accepted journal rankings in a bid to shed light on the adequacy of the ERA classification and ranking of journals. This study revealed that prize-winning scientists tended to exhibit elite publication output for a decade before and a decade after their award, so the analysis of publication patterns focuses on the 88 prize-winners who received their award during 1990-2010, and who are likely to have exhibited elite performance during the ERA assessment period 2005-2010. Subsequent analyses rely on two assumptions about the publication habits of the scientific elite: that they publish a greater proportion of their work in high impact journals, and that they publish in a wide range of journals to reach the most appropriate audience.

It is difficult to establish reliable evidence to test the proposition that experience and acknowledgement (prize-winning) leads to greater participation in more prestigious journals. Part of the difficulty is that of gauging journal prestige across the many facets of forestry. The Journal Impact Factor (Garfield, 2006) is long established and convenient, but many researchers have counselled against its use to appraise research (e.g., Seglen, 1997; Weingart, 2005; Bollen et al, 2009; Vanclay, 2009). The ERA seeks to use its journal ranking as a proxy for the expected future impact of papers published in those journals, which may be best reflected in indicators such as Article Influence (Arendt, 2010; Waltman & van Eck, 2010), and source-normalised impact per paper (SNIP; Moed, 2010). Since SNIP is provided by Scopus, the official data provider to the ERA in 2010, it has been adopted as the benchmark for comparison in this study. Other indicators examined include citation counts, the Impact Factor, Eigenfactor (Bergstrom et al, 2008; West et al, 2010), SCImago Journal Rank (Butler 2008) and h-index (Hirsch, 2005).

**Results and Discussion**

Figure 1 illustrates publication patterns of prize-winning scientists as reflected by SNIP calculated for 2007, chosen to represent the mid-point of the next ERA assessment period (2005-10) and because it is simultaneously recent enough to be current and distant enough to allow reliable assessment of journal impact (Vanclay, 2009). Figure 1a illustrates how the total output of prominent scientists varies over time. This figure is based on the sum of the SNIPs, unadjusted for co-authorship, and may reflect many contributions in 'lowly' journals or fewer contributions in journals with greater impact. One might assume that elite scientists would seek the prestige and wide distribution of journals such as *Science* and *Nature*, but the evidence for this is weak. Analysis of the data in Figure 1 suggests that most prominent scientists increase their impact through coauthorship of a larger number of papers rather than by publishing in journals of higher impact (Figure 1b). This observation is offered non-judgementally, as it is entirely appropriate that prominent scientists attract research students and expand their network of collaboration. This trend is consistent with other research on research productivity of active researchers (e.g., Fox, 1983; Gingras et al, 2008), but the focus on prize-winning scientists is novel as most other research has focused on age and cohort effects in a broader body of scientists (e.g., Lee and Bozeman, 2005; Gonzalez-Brambila and Veloso, 2007; Hall et al, 2007).

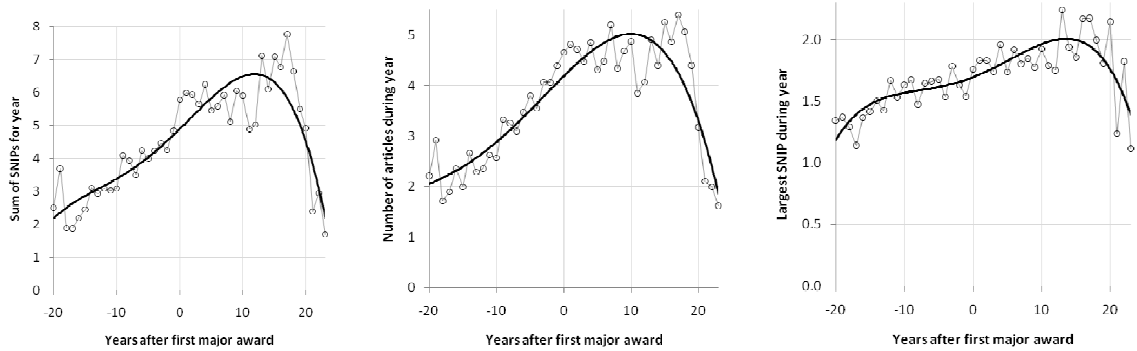

**Figure 1**. Publication activity by elite forest scientists: total output as sum of $SNIP_{2007}$ each year (1a, left), the number of Scopus-listed publications (1b, centre), and the highest $SNIP_{2007}$ of any publication each year (1c right), averaged across all prominent scientists who published in that year. The trend line is a $4^{th}$ order polynomial.

Figure 1c illustrates the maximum SNIP in each year (averaged across all prominent scientists who published that year), showing that there is a slight tendency for prominent scientists to place selected output in 'better' journals as they gain standing in their profession. The annual variations in Figure 1 arise in part because each year reflects a different subset of prominent scientists, as not every scientist published each year. To minimize this sampling effect, Figure 1 is restricted to the 67 scientists who have published in 12 or more years during this period, and the interval -20 to +23 is chosen to ensure that each point represents the average of no fewer than 7 scientists. In the six cases where an author received more than one award, dates are computed from the year of receipt of their first award. It is clear from Figure 1 that prize-winning scientists tend to publish more, and publish better, during the 10-15 years after their award. The same track record is evident for the ten years preceding their award (Figure 1).

Figure 1 draws evidence from the SNIP to support the contention that prominent forest scientists have a high publication impact during the period spanning a decade before and two decades after their award. However, the SNIP (and other metrics) reflects only some aspects of impact, so it is interesting to complement these established metrics, by independently examining the publication patterns of these scientists during these three decades of influence. The intersection of the three decades of influence and the 6-year ERA assessment period 2005-10 means that forest scientists awarded prizes during 1990-2010 are of particular interest. Thus further data analysis is confined to publications during the period relevant to the next ERA assessment (2005-10) by 79 prominent scientists (those amongst the 88 prize-winners during 1990-2010 who have publications visible to Scopus), a total of 1703 publications in 431 journals.

There are many possible indicators of journal standing, but the present analysis confines itself to those reflecting the publication patterns of prominent forest scientists. Four metrics were compared initially: the number of prominent scientists electing to publish in a journal, the number of papers they contribute to each journal, the total citations accruing to those papers, and the citation count adjusted for publication date (obviously, papers published in 2005 may have attracted more citations than papers published in 2010, and citations/year helps to adjust for this temporal effect). These indicators are all correlated, exhibiting a Pearson correlation greater than 0.6 in all cases, but the latter indicator (citation count adjusted for publication date, sum of cites/year) appeared to align most closely with the ERA ranking and other views of journal standing. Figure 2 illustrates this correlation, and reflects the utility of cites/year as a generic indicator.

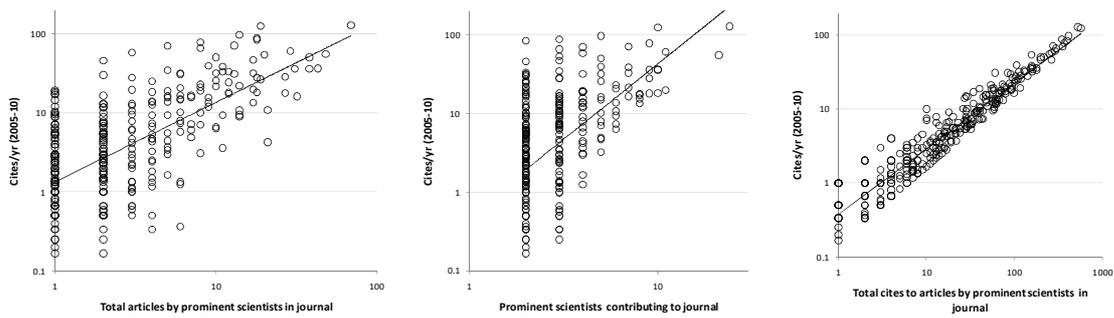

**Figure 2**. Correlations between the four indicators: number of contributors (2a, left), number of contributions (2b, centre), total cites (2c, right) and cites/year (y-axis).

While cites/year is of interest as an indicator, and is related to other indicators derived from the present data (number of contributors and number of contributions), it measures something different to other commonly-accepted indicators such as the ISI Journal Impact Factor (Weingart, 2005), the Article Influence (Waltman & van Eck, 2010), the Scopus SNIP (Moed, 2010), and the h-index (Hirsch, 2005). Figure 3 and Table 1 compare these four indicators with the observed cites/year to prominent forestry scientists, using the Impact Factor (IF), Article Influence (AI), SNIP, and h-indices derived from SCImago (2007), all based on the reference year 2007. While these indicators are clearly correlated, the relatively low correlation suggests that cites/year to elite authors offers an insight complementary to established metrics.

**Table 1**. Correlations between selected indicators of journal impact in 355 journals publishing articles by elite scientists (after applying a logarithm transform).

| Indicator | Cites/yr | IF | AI | SNIP | h-index |
|---|---|---|---|---|---|
| Cites/yr to elite authors | 1 | 0.40 | 0.42 | 0.35 | 0.36 |
| Impact Factor (IF) | 0.40 | 1 | 0.91 | 0.79 | 0.80 |
| Article Influence (AI) | 0.42 | 0.91 | 1 | 0.83 | 0.79 |
| Scopus SNIP | 0.35 | 0.79 | 0.83 | 1 | 0.75 |
| SCImago h-index | 0.36 | 0.80 | 0.79 | 0.75 | 1 |

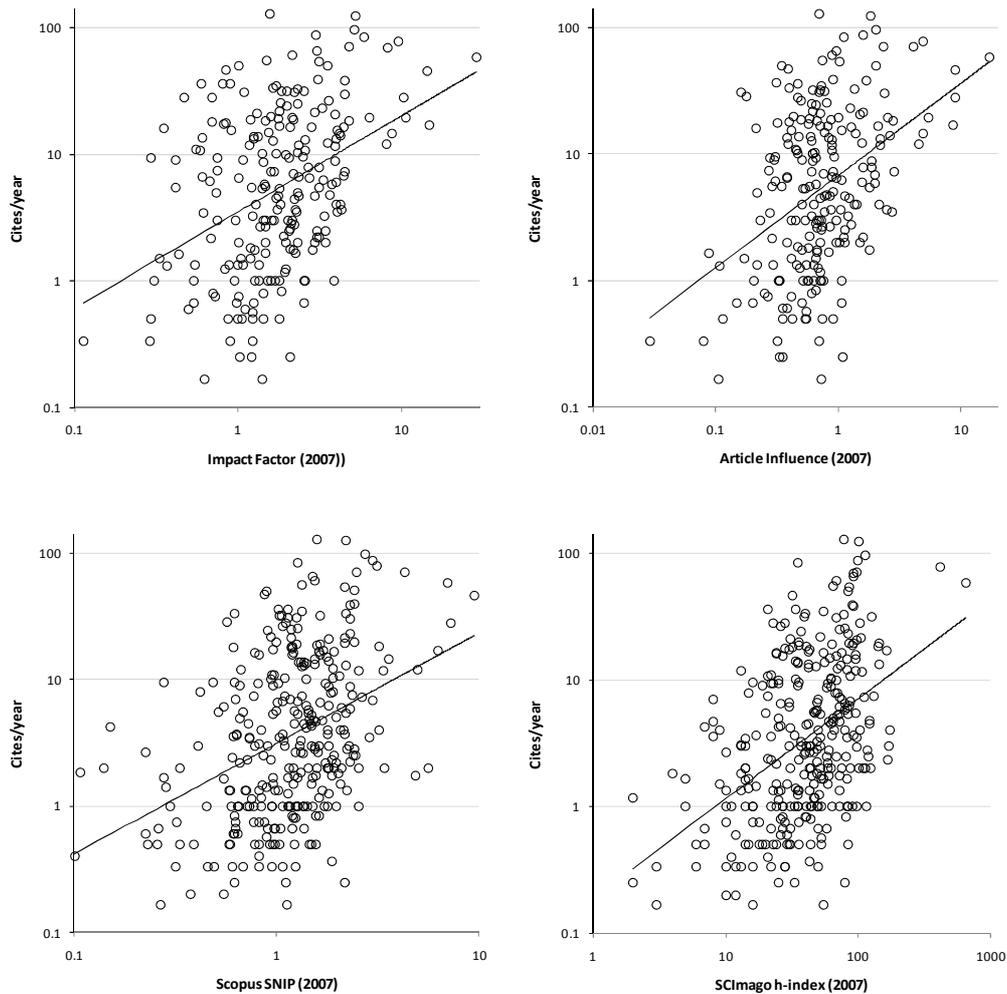

**Figure 3**. Comparison with $IF_{2007}$ (3a, top left), $AI_{2007}$ (3b, top right), $SNIP_{2007}$ (3c, bottom left) and h-index$_{2007}$.(3d bottom right).

These four indicators are summarised for selected journals in Table 2, ranked by cites/year. To enable detailed comparison, Table 2 includes the 'top ten' journals for each indicator: the 10 journals with the greatest number of distinguished contributors, the 10 journals with the greatest number of articles by these contributors, the 10 journals with the largest number of total citations to works by these authors, the 10 journals with the largest number of cites/year, and all eight A-ranked journals classified by ERA as 0705 Forestry Sciences plus the two journals ranked as A* amongst 07 Agricultural and Veterinary Sciences (which includes 0705 Forestry Sciences). Table 2 includes an additional two journals: the *Journal of Vegetation Science* and *Cellulose* which are amongst the top 5% of journals ranked by the ISI Journal Impact Factor within their subject categories Forestry, and Paper and Wood respectively.

**Table 2**. Bibliometric characteristics of selected journals during the period 2005-2010.

| Journal | Total articles | Total cites | Sum of cites/yr | No of contributors | ERA Field of Research (FOR) | ERA Rank |
|---|---|---|---|---|---|---|
| *Forest Ecology and Management* | 69 | 525 | 128 | 24 | 0705 | A |
| *New Phytologist* | 19 | 568 | 124 | 10 | 0605/0607 | A* |
| *Molecular Ecology* | 14 | 415 | 97 | 5 | 0602 | A |
| *Remote Sensing of Environment* | 18 | 344 | 87 | 3 | 0406/0909 | A* |
| *Studies in Mycology* | 18 | 393 | 84 | 2 | -- | -- |
| *PNAS* | 8 | 366 | 78 | 9 | MD | A* |
| *Global Change Biology* | 13 | 275 | 71 | 7 | 05/06 | A* |
| *Ecology Letters* | 5 | 353 | 70 | 4 | 0501/0502/0602 | A* |
| *Biotechnology and Bioengineering* | 8 | 297 | 66 | 3 | 06/09/10 | A |
| *Tree Physiology* | 29 | 288 | 60 | 11 | 0705 | A |
| *Canadian Journal of Forest Research* | 48 | 248 | 55 | 21 | 0705 | A |
| *Journal of Applied Polymer Science* | 38 | 199 | 50 | 5 | 0303/0904/0912 | B |
| *European J. Wood & Wood Products* | 43 | 159 | 36 | 10 | 0705 | B |
| *Holzforschung* | 38 | 153 | 36 | 10 | 0705 | C |
| *Forest Policy and Economics* | 31 | 179 | 36 | 9 | 1402/1605 | C |
| *Agricultural and Forest Meteorology* | 11 | 129 | 33 | 7 | 0401/0705 | A |
| *Australasian Plant Pathology* | 27 | 118 | 28 | 3 | 0605/0607/0703 | C |
| *Wood Science and Technology* | 18 | 141 | 28 | 9 | 0607/0705/0912 | B |
| *Annals of Forest Science* | 17 | 76 | 20 | 11 | 0705 | B |
| *Forestry Chronicle* | 27 | 70 | 18 | 8 | 0705 | C |
| *Scandinavian J. Forest Research* | 18 | 58 | 18 | 9 | 0705 | B |
| *Trees – Structure and Function* | 12 | 92 | 18 | 10 | 0705 | B |
| *Forest Products Journal* | 32 | 79 | 16 | 5 | 0705 | C |
| *Forestry* | 9 | 47 | 16 | 8 | 0705 | A |
| *Forest Science* | 17 | 68 | 13 | 8 | 0705 | A |
| *Tree Genetics and Genomes* | 9 | 34 | 12 | 4 | 0604/0705/1001 | A |
| *Conservation Biology* | 2 | 62 | 12 | 3 | 05/06/07 | A* |
| *Cellulose* | 6 | 26 | 10 | 6 | 0303/0912 | B |
| *Journal of Vegetation Science* | 1 | 11 | 4 | 2 | 0607 | B |
| *Applied & Environmental Microbiology* | 1 | 4 | 4 | 2 | 06/07/10 | A* |
| *International Journal of Wildland Fire* | 0 | 0 | 0 | 0 | 0705 | A |

Although the various indicators in Table 2 are highly correlated, there are some notable outliers. *Forest Policy and Economics* has become prominent as a publication outlet for elite scientists (9) and carries a relatively large number of contributions (31), but these are cited rather infrequently (179 times in total or 36 cites/year). *Studies in Mycology* has few contributors (2) for a rather high-impact journal (84 cites/year), reflecting the narrow focus of the journal. And the A-ranked journal *International Journal of Wildland Fire* has received no attention from the prize-winning scientists considered in this paper, suggesting that either few awards have been made for fire research, and/or that this journal is misclassified. Finally, notwithstanding the other indicators, journals appear to be ranked A* only if they are not classified as 0705 Forestry. This discrepancy is further examined in Figure 4.

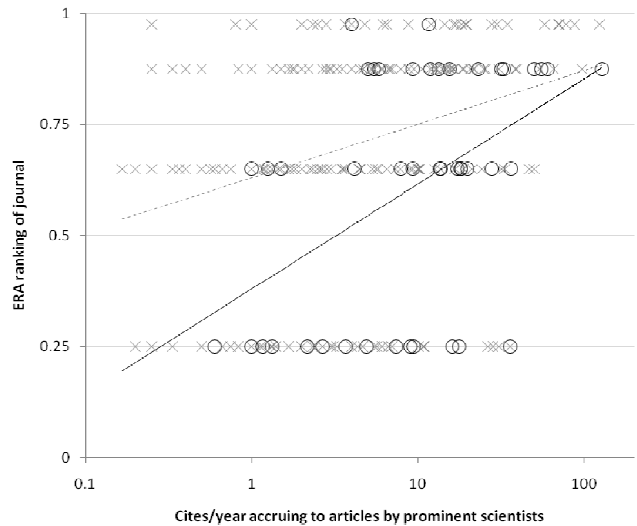

**Figure 4**. Citation patterns accruing to journals ranked by ERA as forestry (0705 Forestry and 07 Agriculture, shown as circles, trend as solid line) and non-forestry (shown as crosses, trend dotted).

Figure 4 shows citations/year accruing to work published during 2005-2010 by the prize-winning forestry scientists identified in this study, within each of the four journal rankings defined by ERA in 2010. The two trend lines for forestry (i.e., journals classified as FOR 07 and 0705) and non-forestry journals (other FOR codes, including multidisciplinary) have very different slopes, so for instance, work published in B-ranked forestry journals tends to accrue about ten times as many citations as work by the same scientists published in B-ranked 'non-forestry' journals. There are two ways to interpret these differing trends. One interpretation is that forestry articles published in non-forestry journals attract fewer citations than typical for the journal because such articles are seen by a disinterested audience. While this situation may occur occasionally, the present study draws on work by elite prize-winners who are unlikely to hide their output in obscure journals. It also overlooks the role of informational retrieval systems such as Scopus, Web of Science and Google Scholar that rely on keywords rather than journal subscriptions. An alternative interpretation of these two trends is that forestry journals are ranked lower by ERA than journals of comparable impact in other fields. Either interpretation leads to the need for a reliable and equitable ranking of journals within each FOR.

In Figure 4, the trend for forestry journals (solid line) is close to diagonal, consistent with the assumptions of the ERA ranking in assuming a 5:15:30:50 distribution amongst A*, A, B and C-ranked journals, which is surprising given that the work under examination is by the elite amongst forestry researchers during their prime. The dotted trend reveals that work by the same authors is ranked more highly by ERA (for the same citation impact) when published in non-forestry journals. Figure 4 suggests that there may be inadequacies in the 2010 ERA ranking that should be addresses in the pending revision (Atkinson and McBeath, 2010).

The award-winning forestry scientists published in 424 journals during 2005-2010, but many of these journals carried only one or two articles by prominent scientists. The top 25 journals (Table 3) carried approximately 30% of the articles and accrued about 50% of the citations, so these are the journals that should be examined more closely. The stated intention of the ERA was that the top 5% of journals should be ranked A*, so it is appropriate to consider the ERA ranking of these journals (Table 3), since these 5% of journals favoured by the elite amongst researchers would seem likely candidates for A* ranking.

**Table 3**. Top 25 most-frequently cited journals in which elite forest scientists choose to publish.

| Journal | Articles | Total Cites | Cites per year | Rank Other FOR | Rank Multi-FOR | Rank 0705 |
|---|---|---|---|---|---|---|
| *Forest Ecology and Management* | 69 | 525 | 128 | | | A |
| *New Phytologist* | 19 | 568 | 124 | A* | | |
| *Molecular Ecology* | 14 | 415 | 97 | A | | |
| *Remote Sensing of Environment* | 18 | 344 | 88 | A* | | |
| *Studies in Mycology* | 18 | 393 | 84 | -- | | |
| *PNAS* | 8 | 366 | 78 | | A* | |
| *Global Change Biology* | 13 | 275 | 71 | A* | | |
| *Ecology Letters* | 5 | 353 | 70 | A* | | |
| *Biotechnology and Bioengineering* | 8 | 297 | 66 | A | | |
| *Tree Physiology* | 29 | 288 | 60 | | | A |
| *Nature* | 3 | 266 | 58 | | A* | |
| *Canadian Journal of Forest Research* | 48 | 248 | 55 | | | A |
| Median of top 2.5% | 16 | 348 | 75 | A* | A* | A |
| *Environmental Pollution* | 20 | 156 | 54 | | A | |
| *Ecological Applications* | 10 | 215 | 50 | | A | |
| *Journal of Applied Polymer Science* | 38 | 199 | 50 | B | | |
| *J. Adhesion Science and Technology* | 17 | 199 | 47 | B | | |
| *Materials Science & Engineering Reports* | 2 | 265 | 46 | A* | | |
| *Bioresource Technology* | 9 | 106 | 39 | | A | |
| *Plant, Cell and Environment* | 11 | 157 | 38 | A | | |
| *European J. Wood and Wood Products* | 43 | 159 | 36 | | | B |
| *Holzforschung* | 38 | 153 | 36 | | | C |
| *Forest Policy and Economics* | 31 | 179 | 36 | C | | |
| *Chemical Engineering Journal* | 5 | 161 | 35 | A* | | |
| *Applied Biochemistry and Biotechnology* | 12 | 179 | 33 | B | | |
| *Agricultural and Forest Meteorology* | 11 | 129 | 33 | | A | |
| Median of top 5% | 14 | 248 | 54 | A | A | A |

Table 3 provides further insights into weaknesses of the ERA classification and ranking. The journal *Studies in Mycology* appears to have been overlooked from the classification. There are a large number of A*-ranked journals in the 'non-forestry' column, but none in the 'forestry' column of Table 3, despite journals of apparently comparable standing, suggesting an apparent bias against forestry in the journal rankings. It is somewhat surprising that so few of these journals are ranked A*, since by one yardstick they represent the top 5% of journals frequented by elite scientists at their peak performance. In addition, there are a large number of journals in which prominent forest scientists publish, that are not classified 0705 Forestry Sciences, suggesting the need for more multiple classifications amongst these journals.

**Conclusion**

Table 3 offers a compelling argument that the classification and ranking of journals in 0705 Forestry Sciences warrants further consideration. There appears to be a strong case to rank as A* at least three journals, including *Forest Ecology and Management*, *Tree Physiology* and *Canadian Journal of Forest Research*. There is also a strong case to add additional classifications for several journals not currently classified as 0705 Forestry Sciences. These findings are consistent with other studies drawing on different sources of data.